\documentclass[eqsecnum,aps,preprint,epsf]{revtex4}
\usepackage{graphicx}
\usepackage{wrapfig}
\begin{document}

\title{Antiproton Evolution in Little Bangs and Big Bang}

\author{H. Schade}
\author{ B. K\"ampfer}
\affiliation{Forschungszentrum Dresden-Rossendorf, PF 510119, 01314 Dresden, Germany\\
TU Dresden, Institut f\"ur Theoretische Physik 01062 Dresden, Germany}


\begin{abstract}
The abundances of antiprotons and protons are considered
within momentum-integrated Boltzmann equations
describing Little Bangs, i.e., 
fireballs created in relativistic heavy-ion collisions.
Despite of a large antiproton annihilation cross section
we find a small drop of the ratio of antiprotons to protons
from 170 MeV (chemical freeze-out temperature)
until 100 MeV (kinetic freeze-out temperature) for CERN-SPS and BNL-RHIC energies
thus corroborating the solution of the previously exposed ''antiproton puzzle''.
In contrast, the Big Bang evolves so slowly that the antibaryons
are kept for a long time in equilibrium resulting in an exceedingly
small fraction. The adiabatic path of cosmic matter 
in the phase diagram of strongly interacting matter is mapped out.
 
\end{abstract}
\maketitle

\section{Introduction}

The abundances of hadrons emerging from relativistic heavy-ion collisions
can be described with surprisingly high accuracy by a thermo-statistical
model \cite{hadro_chemistry}. Over a wide range of bombarding energies
essentially two parameters, the temperature $T_{chem}$ and the baryo-chemical
potential $\mu_{chem}$, adjusted to experimental data give a smooth curve
$T_{chem}(\mu_{chem})$, the so-called chemical freeze-out curve,
being an important landmark in the phase diagram of strongly interacting matter.
At a given beam energy
the variation of $T_{chem}, \, \mu_{chem}$ for various centralities is fairly small 
\cite{our_PRC}. 

The transverse momentum spectra of various hadron species, at a given beam energy,
may also be described by a set of parameters $T_{kin}$, $\mu_{kin}$, where additionally
a flow parameter may be employed. It was argued, e.g.\ in \cite{bk_early}, and later
confirmed, e.g.\ in \cite{mT_parameters}, that $T_{chem} > T_{kin}$ holds
in central collisions at high energies.
This is often interpreted in a schematic picture as ceasing of inelastic (chemical)
reactions at $T_{chem}$, while in a later stage of the fireball evolution a kinetic
freeze-out happens at $T_{kin}$ where the elastic interactions are no longer efficient
enough to modify the momentum distributions.
A different view was advocated in \cite{BB} with $T_{chem} \approx T _{kin}$.
In the former scenario there is still ''hadronic life'' after freezing out 
the chemical composition, while the latter scenario would mean a sudden
termination of inelastic and elastic reactions roughly at the same time.

A possible problem in the commonly accepted first scenario with
$T_{chem} > T_{kin}$ was pointed out in \cite{antip_puzzle_Shuryak}:
The antiproton ($\bar p$) annihilation cross section is so large
that the $\bar p$ survival until kinetic freeze-out needs special consideration.
This was exposed as the ''antiproton puzzle''.
In \cite{antip_puzzle_Shuryak,antip_puzzle_Rapp} it was shown
that chemical off-equilibrium combined with
multi-hadron reactions give rise to sufficient antiproton regeneration
ensuring a moderate drop of the ratio of abundances $\bar p/p$
from $T_{chem}$ until $T_{kin}$ for CERN-SPS energies. This fits well in the
above first hadro-chemical scenario \cite{hadro_chem_Rapp}.

It is the aim of this note to reconsider the off-equilibrium evolution of antiprotons
in an expanding hadron fireball. We extend the consideration of 
\cite{antip_puzzle_Rapp,hadro_chem_Rapp}
towards BNL-RHIC and CERN-LHC top energies. In a schematic model we show that the
$\bar p / p$ ratio at CERN-SPS energies decreases from a given temperature
$T_{chem} \approx 170$ MeV until kinetic freeze-out temperature 
$T_{kin} \approx 100 - 120$ MeV
only by a small fraction, while at BLN-RHIC energies the variation
of $\bar p / p$ in such a temperature interval is even negligible.
We expect that at CERN-LHC energies the latter statement also applies.  
We employ here a transparent (thus simplified) model to elaborate
the features of the evolution towards off-equilibrium
after an assumed chemical equilibrium
which complies with the mentioned hadro-chemical model \cite{hadro_chemistry}.
In contrast to this, the involved and sophisticated transport models
give a much more complicated description, both for hadro-chemistry in general
\cite{Stocker} and for the antiprotons especially \cite{Cassing,Bleicher,Pang}.

Furthermore, we emphasize similarities and differences to antimatter (baryon)
evolution in the early universe after confinement.
The adiabatic path of cosmic baryon matter is exhibited in the
phase diagram of strongly interacting matter.  

The paper is organized as follows. In Section 2 we discuss the appropriate
evolution equations for protons (baryons) and antiprotons (antibaryons)
coupled by a conservation law.
The main emphasis is devoted to applications in relativistic heavy-ion
collisions (Section 3) covering situations with 
baryons and antibaryons being nearly symmetric
or very asymmetric. Section 4 describes the cosmic baryon matter.
The summary can be found in Section 5. Appendix A sketches the derivation of the used
evolution equations.

\section{Off-Equilibrium Evolution of Antiprotons}

Our starting point is the pair of momentum-integrated Boltzmann equations
(see Appendix A)
\begin{eqnarray}
\frac{d Y_+}{d x} &=& - \frac{\Lambda_{(\xi)}}{x^\xi} \left( Y_+ (Y_+ - \eta) - Y_{eq}^2 \right), \label{eq1}\\
\frac{d Y_-}{d x} &=& - \frac{\Lambda_{(\xi)}}{x^\xi} \left( Y_- (Y_- + \eta) - Y_{eq}^2 \right), \label{eq2}
\end{eqnarray}
where $Y_\pm = n_\pm / s$ are the baryon densities
$n_\pm$ normalized to entropy density $s = \frac{2 \pi^2}{45} h_{eff} T^3$.
We attribute $Y_+, n_+$ to protons ($p$) and $Y_-, n_-$ to antiprotons ($\bar p$).
$h_{eff}$ is the effective number of degrees of freedom of hadrons
in the fireball with comoving volume $V(t)$. Here, we assume a spatially homogeneous
fireball and apply for protons and antiprotons the Boltzmann approximation,
which is appropriate for the following since $x \equiv m_N /T$ is larger than unity. Therefore,
$Y_{eq} = \frac{45}{4 \pi^4} \frac{2}{h_{eff}} x^2 K_2(x)$ with
the Bessel function $K_2$. 
Both evolution equations 
refer to pair-wise annihilations and regenerations, thus
$Y_+ - Y_- \equiv \eta = const$,
i.e.\ $\dot Y_+ = \dot Y_-$. 
The two parameters in the evolution equations, $\Lambda_{(\xi)}$
and $\eta$, determine together with initial conditions
the off-equilibrium dynamics. In what follows
we are going to explore their interplay for Little Bangs and Big Bang. 

For Little Bangs ($\xi = 4$, see Appendix A) one has  
\begin{equation}
\Lambda_{(4)} = 3 \langle \sigma v \rangle \bar \tau m_N^3 
\frac{2 \pi^2}{45} h_{eff}
\left(1 + \tau \frac{\dot h_{eff}}{ h_{eff}} \right)^{-1}.
\end{equation}
It encodes essentially the thermally averaged annihilation cross section 
$\langle \sigma v \rangle$, and
$\dot h_{eff}$ accounts for the time variation of the effective degrees
of freedom.
In deriving Eqs.~(\ref{eq1}, \ref{eq2})
for Little Bangs
we use adiabaticity of the expanding fireball, i.e.\
$s V = const$. 
Instead of a specific expansion model we utilize $V / \dot V = \bar \tau$
with $\bar \tau$ as characteristic time scale.
This allows us to formulate
the evolution as a function of $x$ instead of a function of time.

In the Big Bang ($\xi = 2$, see Appendix A and \cite{Turner,Gondolo,Griest})
the dimensionless factor $\Lambda_{(\xi)}$ reads
\begin{equation}
\Lambda_{(2)} = \langle \sigma v \rangle \,g_*^{1/2} \,M_{Pl} \,m_N \,
\sqrt{\frac{\pi}{45}}
\end{equation}
with $g_*^{1/2} = h_{eff} \, g_{eff}^{- 1/2} 
\left(1 + \displaystyle\frac{T}{3 h_{eff}} 
\displaystyle\frac{\dot{h}_{eff}}{\dot{T}} \right)$ and
$g_{eff}$ determining the effective degrees of freedom relevant
for the energy density, i.e.\ $e= \frac{\pi^2}{30} g_{eff} \, T^4$, 
and $M_{Pl}$ as Planck mass and $m_N$ again as nucleon mass.
Clearly, different $h_{eff}$'s apply in Little Bangs and Big Bang.  

Using $\langle \sigma v \rangle = C / m_\pi^2$
with $C = {\cal O}(1)$ and with $m_\pi$ denoting the pion ($\pi^0$) mass one arrives for
$\bar\tau \sim 5$ fm/c at $\Lambda_{(4)} \sim {\cal O}(10^4)$ as an estimate
(for $h_{eff}$ see below), while
$\Lambda_{(2)} \sim {\cal O}(10^{21})$ for $g_*^{1/2} \sim 4$ \cite{Hindmarsh_PoS}
highlights the vast difference of Big Bang and Little Bang dynamics.
Actually, the thermally averaged cross section is
$
\langle \sigma v \rangle =
\frac{\int_{2x}^\infty d \xi \xi^2 (\xi^2 - 4 x^2) K_1(\xi) \sigma (p_{lab})}{4 x^4 K_2^2(x)}
$
\cite{Gondolo} with $p_{lab} = T \xi \sqrt{\xi^2 - 4 x^2} /(2 x)$,
see also \cite{Kapusta_Mekian}.
Employing $\sigma (p_{lab}) = (40 p_{lab, GeV/c}^{-0.5} + 24 p_{lab,GeV/c}^{-1.1})$ mb 
\protect\cite{antip_puzzle_Rapp}
or $(38 + 35 p_{lab, GeV/c}^{-1})$ mb  
for the $\bar p p$ annihilation cross section
one arrives at
$\langle \sigma v \rangle \simeq$ 42.5 - 47 mb in the temperature interval 170 - 100 MeV
or 51.5 mb fairly independent of $T$, thus yielding $C \sim 2$.

\section{Results for Little Bangs}

Equations (\ref{eq1}, \ref{eq2}) are of Riccati type
with no general analytical
solution in closed form. The solutions, to be found numerically,
depend on $\Lambda_{(\xi)}$ and the initial
conditions (encoded in $\eta$). For the latter ones we employ 
$T_{chem} = T_0 = 170$ MeV
and $\mu_{chem} = \mu_0 = 250$ MeV (SPS top energy resulting in $n_- / n_+ = 0.052$)
or  $\mu_{chem} = \mu_0 = 25$  MeV (RHIC top energy resulting in $n_- / n_+ = 0.75$),
thus neglecting a possible small change of chemical freeze-out temperature
when going from SPS to RHIC
but catching the typical values extracted in the data analysis 
\cite{hadro_chemistry,our_PRC}. 
SPS energies are in the realm $\mu_{chem} > T_{chem}$, while RHIC and
LHC operate in the region $\mu_{chem} < T_{chem}$.
A good approximation of $h_{eff}$ is provided by 
$h_{eff}(T)=h_1+\hat b (T-T_1)+ \hat c (T-T_1)^2$
with 
$\hat b=\left[h_2-h_1-(h_3-h_1) (T_2-T_1)^2/(T_3-T_1)^2\right]/\left[T_2-T_1-(T_2-T_1)^2/(T_3-T_1)\right]$ and
$\hat c=(h_3-h_1)/(T_3-T_1)^2-\hat b/(T_3-T_1)$
and $(T_1,T_2,T_3,h_1,h_2,h_3) = (0.100,0.150,0.175,3.0,9.8,17.5)$
for small  chemical potentials as appropriate for RHIC top energies
and $(0.100,0.150,0.175,5.411,13.150,21.750)$ for an isentrope with entropy
per baryon equal to 23 as appropriate for SPS top energy. These values are for
the resonance gas model with the first hundred hadronic states as used
in \cite{hadro_chemistry} and for temperatures in GeV. While the scales 
$\langle \sigma v \rangle$ and $\bar \tau$ as well as $h_{eff}$ 
are condensed into one dimensionless
parameter $\Lambda_{(4)}$ which would be useful for $h_{eff} = const$, the strong variation
of $h_{eff}$ in the considered temperature range along isentropic trajectories 
makes $\Lambda_{(4)}$ not a concise characteristic quantity. 
Therefore, we present the results for various values
of $C$ to expose the influence of annihilations and to discriminate them
from diminishing densities due to expansion.  

\begin{figure}[h]  
\vskip -12mm
\includegraphics[width=0.6\textwidth]{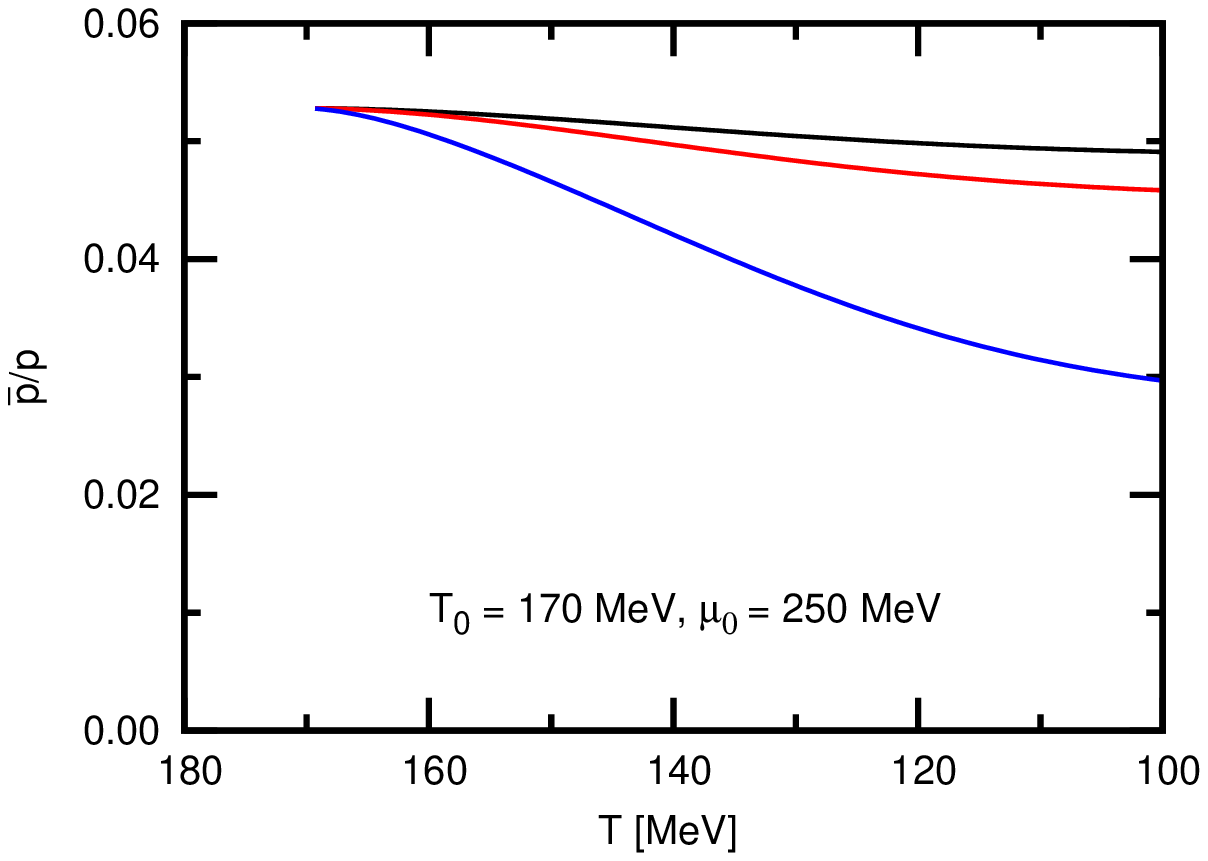} %
\caption{\label{fig1} (Color online)
The ratio of antiprotons to protons
as a function of temperature for various values of $C = 1$, 2, 10 (from top to bottom).
Initial conditions for SPS as described in the text. }
\end{figure}

In Fig.~\ref{fig1} the change of $n_- / n_+$ 
(denoted by $\bar p / p$) as a function of the temperature
is exhibited. We note that, similar to the consideration in \cite{antip_puzzle_Rapp},
the ratio drops only by a small fraction even when extending
the evolution down to a temperature of 100 MeV. The reason is the small change
of $Y_\pm$. 
The experimental mid-rapidity value of $\bar p / p$ is $0.058 \pm 0.005$ 
(statistical error) in central collisions
of Pb + Pb at $E_{beam} = 158$ AGeV \cite{NA49}. Clearly, our above mentioned
initial value is already somewhat below this experimental value. If one would 
change $\mu_0$ to 235 MeV, being still in the range of admissible values
according to \cite{hadro_chemistry}, the start value would be $n_- / n_+ = 0.063$
and all curves exhibited in Fig.~\ref{fig1} are then upshifted, roughly by factor 1.2.
As a consequence, values of $C < 4$
would be compatible with the data. While this range of $C$
is realistic, such a fine
tuning is not appropriate given the schematic character of our model
(use of a characteristic time scale $\bar \tau$ for the expanding fireball,
no feeding etc.). This refrains us from considering further details like
the centrality dependence. Nevertheless, one may consider the beam energy
systematics for central collisions. Instead of individually selected
values of $\mu_{chem}$ and $T_{chem}$, one may use a global fit 
of many hadron ratios with the parametrization \cite{hadro_chemistry}, 
$\mu_{chem} = 1303\,\mbox{MeV}\, /(1 + 0.286 \sqrt{s_{NN}}/\mbox{GeV})$
and $T_{chem} = 162\, \mbox{MeV}\,\left(1 - (0.7+\exp[(\sqrt{s_{NN}}/\mbox{GeV}-2.9)/1.5])^{-1}\right)$
(where $\sqrt{s_{NN}}$ is in GeV)
and get initial values for $n_- / n_+$ being
15\%, 1.5\%, 18\%, 30\% and 29\% larger than the experimental ratios $\bar p/p$
quoted in \cite{NA49} for central Pb + Pb collisions at SPS beam energies
of 158, 80, 40, 30 and 20 AGeV. The offset of $n_- / n_+$
above the experimental $\bar p/p$ value \cite{PHENIX}
at $\sqrt{s_{NN}} = 200$ GeV is 36\% according to this parametrization.
With the exception of the data situation for beam energy of 80 AGeV, 
there is room for a 20\% drop, in average, of $\bar p/p$ towards kinetic
freeze-out, consistent with the results in Fig.~\ref{fig1}.

\begin{figure}[h]  
\vskip -6mm
\includegraphics[width=0.67\textwidth]{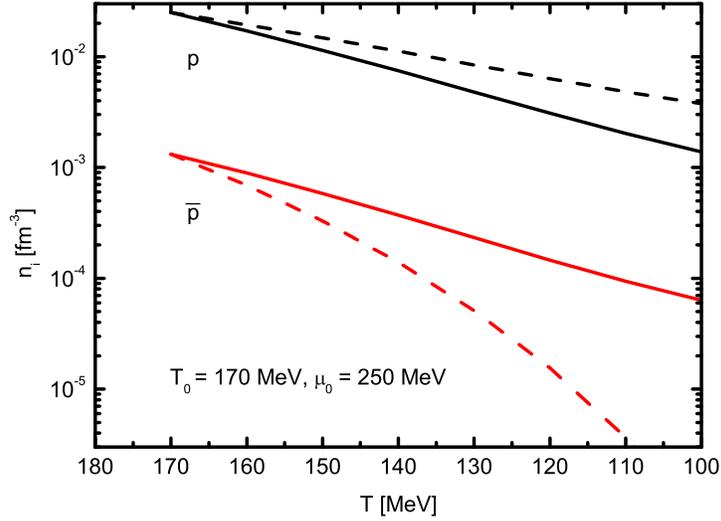} %
\vskip -6mm
\caption{\label{fig2} (Color online)
The proton density and antiproton density
as a function of temperature. Initial conditions for SPS.
Solid curves: actual densities $n_\pm$ for the off-equilibrium evolution
(i.e.\ solutions of Eqs.~(\protect\ref{eq1}, \protect\ref{eq2})), 
dashed curves: fiducial equilibrium densities $n_\pm^{eq}$
which need special explanation (see text). For $C = 2$.}
\end{figure}

In contrast to the ratio $\bar p / p$, the densities rapidly change with dropping
temperature, as exhibited in Fig.~\ref{fig2}. A notable point is
the increasingly strong departure of $n_-$ from the chemical equilibrium 
value $n_-^{eq}$ towards kinetic freeze-out.
For this comparison we have determined $n_\pm^{eq}$ by
$s Y_{eq} \exp(\pm \mu/T)$ with $\mu$ from the resonance gas model
along the isentropic curve imposing baryon conservation (a convenient
parametrization is provided by  
$\mu = \tilde a / T^{\tilde x} + \tilde b$ with
$(\tilde x, \tilde a, \tilde b) = (1, 0.045198563, -0.015873836)$ --
$T$ and $\mu$ in units of GeV).
The emerging relation $n_+^{eq} > n_+$ (see Fig.~\ref{fig2}) is surprising
at the first glance. In fact, baryon conservation enforces 
$n_N > n_N^{(0)} V(t=0) / V(t > 0)$ for the net nucleon density.
Since the antibaryon density in the situation at hand is significantly
smaller than the baryon density, the nucleon density also decreases
slower than $V(t=0) / V(t > 0)$. As our evolution equations impose
pair-wise annihilations and regenerations, for $n_+ \gg n_-$,
$n_+$ goes approximately with $V(t=0) / V(t)$, while the fiducial
density obeys $n_+^{eq} > n_+^{(0)} V(t=0) / V(t > 0)$. This ostensible
ambiguity can be resolved by introducing effective chemical potentials
in line with \cite{antip_puzzle_Shuryak,antip_puzzle_Rapp,hadro_chem_Rapp}.
This, however, is not necessary for the present purposes, as the
combination $n_+^{eq} n_-^{eq}$ (or $Y_+^{eq} Y_-^{eq}$) enter the evolution
equations, and neither $n_+^{eq}$ nor $n_-^{eq}$ separately:
In the employed Boltzmann approximation, the chemical potentials
cancel in $Y_+^{eq} Y_-^{eq} = Y_{eq}^2$. Discarding this gain
(or recombination) term would result in a significantly reduced ratio $\bar p / p$,
thus emphasizing the importance of the back-reaction, as already stressed
in \cite{antip_puzzle_Shuryak,antip_puzzle_Rapp,hadro_chem_Rapp}.
(In \cite{antip_puzzle_Rapp} the regeneration term is of utmost importance
to counteract the stronger annihilations at the total baryon content.)    

Summarizing, the
inspection of Fig.~\ref{fig2} reveals the strong deviation from
equilibrium, i.e., despite of the large annihilation cross section
the expansion is too rapid to maintain chemical equilibrium of antiprotons.
For larger values of $\langle \sigma v \rangle$ parameterized by $C$,
$n_-$ follows more closely the fiducial density
$n_-^{eq}$ and, as a consequence, the antiproton to proton ratio
drops stronger during cooling (see also Fig.~\ref{fig1}).
A larger value of the expansion time scale $\bar \tau$ reduces also
the ratio $\bar p / p$: For $\bar \tau = 10$ fm/c we get 0.040 at $T = 100$ MeV.
In contrast, a shorter expansion time scale keeps the ratio
at higher values, say 0.048 for $\bar \tau = 3$ fm/c. 
As $n_+$ is essentially not modified, one can infer the
corresponding values of $n_-$ from these numbers. 

\begin{figure}[h]  %
\includegraphics[width=0.6\textwidth]{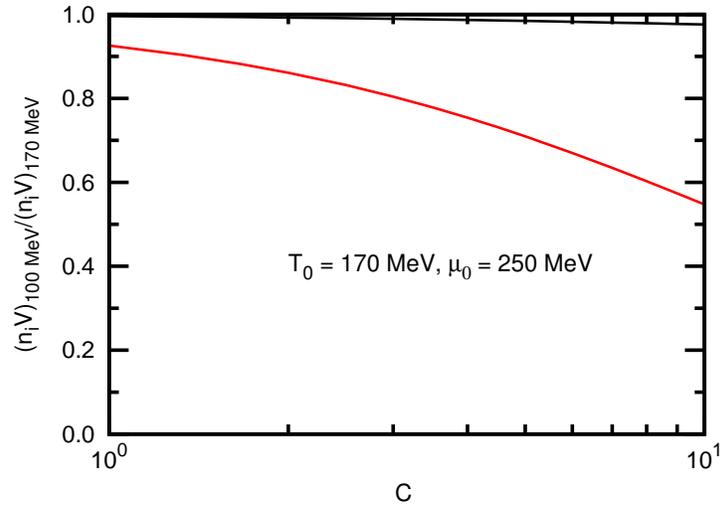} %
\caption{\label{fig2a} (Color online)
The proton yield (upper curve) and antiproton yield
(lower curve) at 100 MeV normalized to those at 170 MeV
as a function of $C$.
Initial conditions for SPS.}
\end{figure}

Reference \cite{hadro_chemistry} (first quotation) states that the thermo-statistical model
applies not only to selected ratios of hadron yields but also to yields themselves.
It is instructive, therefore, to consider the evolution of the yields
from chemical freeze-out towards kinetic freeze-out temperatures.
We normalize the yields according to
$(n_i V)_{100 \,MeV} / (n_i V)_{170 \,MeV}$ and exhibit in Fig.~\ref{fig2a}
the dependence on $C = \langle \sigma v \rangle / m_\pi^2$.
The yield of antiprotons depends smoothly on $C$; values $C > 3$
would cause a sizeable annihilation 
which would obscure the hadro-chemistry picture.  

\begin{figure}[h]  %
\vskip 15mm    
\includegraphics[width=0.63\textwidth]{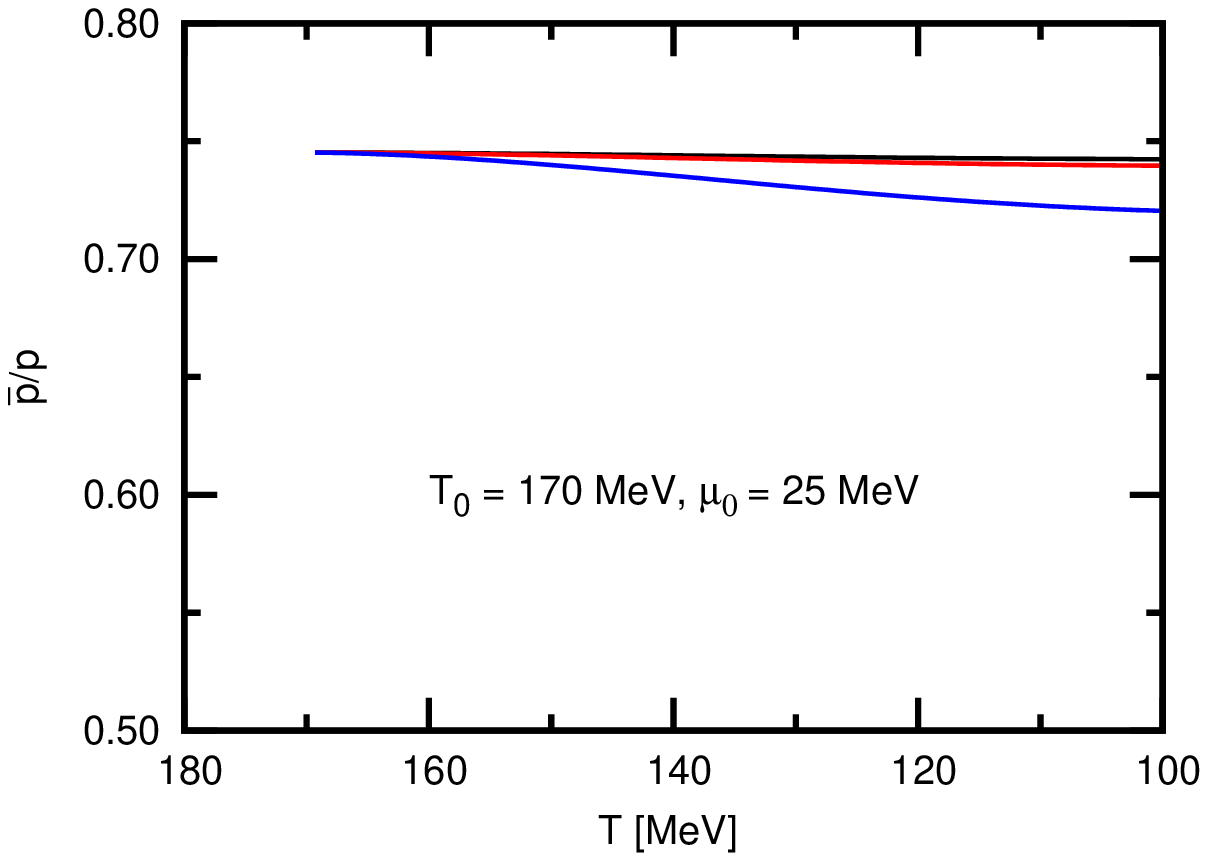} %
\caption{\label{fig3} (Color online)
As Fig.~\ref{fig1} but for RHIC initial conditions. }
\end{figure}

\begin{figure}[h]  %
\includegraphics[width=0.67\textwidth]{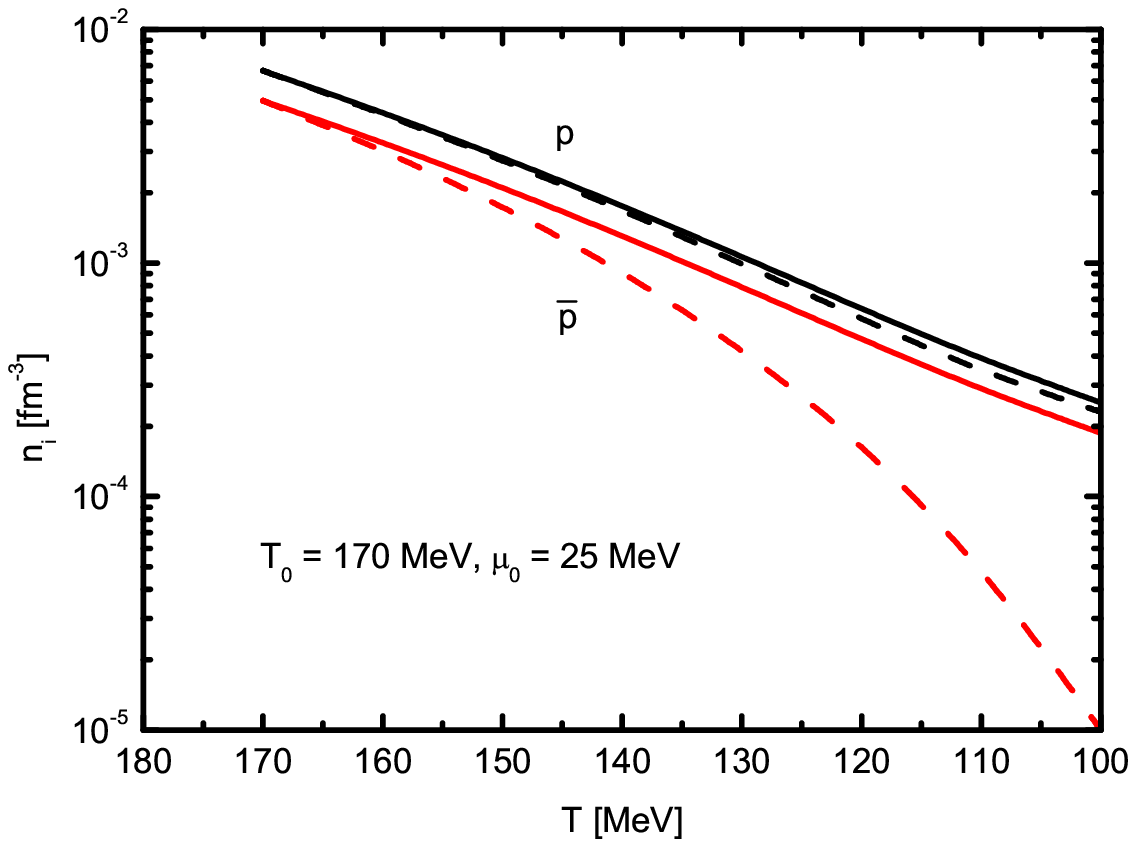} %
\vskip -6mm
\caption{\label{fig4} (Color online)
As Fig.~\ref{fig2} but for RHIC initial conditions.  }
\end{figure}

\begin{figure}[h]  %
\includegraphics[width=0.63\textwidth]{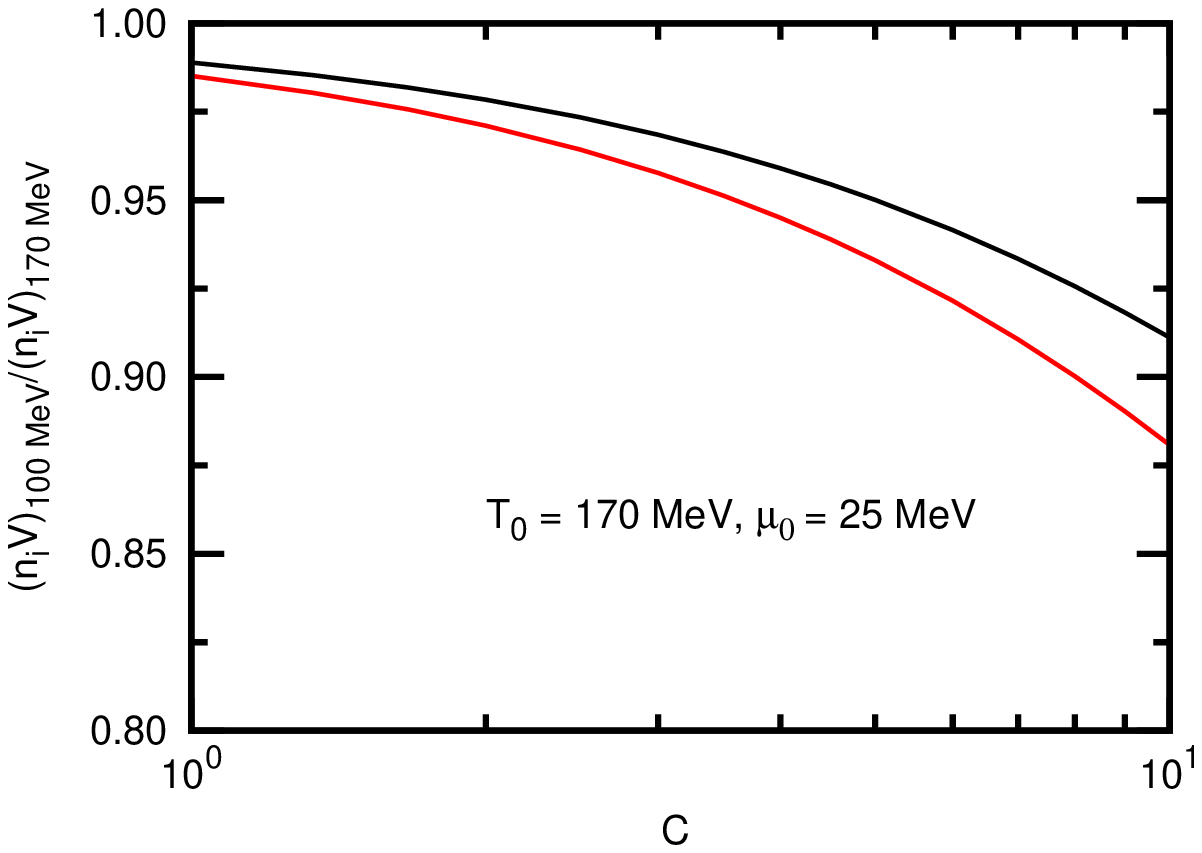} %
\caption{\label{fig4a} (Color online)
As Fig.~\ref{fig2a} but for RHIC initial conditions.  }
\end{figure}

For RHIC energy the same features hold. However, different initial conditions
cause some different evolution: The ratio $\bar p / p$ is fairly insensitive
to $C$, see Fig.~\ref{fig3}. The departure from equilibrium for protons
is nearly as strong as for antiprotons, as shown in Fig.~\ref{fig4}.
(Here, the above parametrization of $\mu(T)$ with
$( \tilde x, \tilde a, \tilde b) = (4.8, 2.2671929 \times 10^{-6}, 0.013797031)$
applies.)
Also the yields in Fig.~\ref{fig4a}
drop with increasing values of $C$ as they are more dragged by
the respective equilibrium values. For $C < 10$
the reduction via annihilation is less than 10\% thus not invalidating the consistency
of hadro-chemistry with a late kinetic freeze-out. 
Experimentally, one finds $\bar p / p = 0.731 \pm 0.011 \pm 0.062$ \cite{PHENIX}
in central collisions Au + Au at $\sqrt{s_{NN}} = 200$ GeV which is fairly independent
of centrality. This value compares well with the results in Fig.~\ref{fig3},
even for large $C$. 

For smaller values of $\mu_{chem}$ the difference of $n_-$ and $n_+$
becomes smaller: Annihilation diminishes both $n_-$ and $n_+$ by the
same amount thus keeping the ratio $n_- / n_+$ nearly constant.
This consideration applies in particular for LHC.
In contrast, for the above
baryon-antibaryon asymmetric situation at SPS, given by a larger value
of $\mu_{chem}$, the annihilation of $p$ by the small admixture of $\bar p$
is not severe, and only the evolution equation for $n_-$ is sufficient,
as exploited in \cite{antip_puzzle_Rapp}.

Resolving the ''antiproton  puzzle'' means demonstrating that
the $\bar p / p$ ratio does not change noticeably from
$T_{chem}$ until $T_{kin}$. At high beam energies (say, for RHIC and 
LHC energies) corresponding to smaller values of
the normalized particle-antiparticle number difference $\eta$ 
this seems to be fairly robust.
At SPS one sees already a sensitivity to the interplay of thermally averaged
annihilation cross section and expansion dynamics. 
At AGS and later on planned FAIR energies one expects a stronger
drop of the $\bar p / p$ ratio. However, the differences of
$T_{chem}$ and $T_{kin}$ may be much smaller so that again
annihilation is less severe.

Our results base on a few assumptions which we recollect here:
(i) kinetic equilibrium (which is left at the kinetic freeze-out, 
see \cite{Csernai} for dealing with the freeze-out itself);
(ii) spatial homogeneity;
(iii) expansion dynamics characterized by one time scale;
(iv) restriction to one hadron species, $p, \bar p$ (the other hadrons
are implicitly in the heat bath, encoded in $h_{eff}$ \footnote{
A more detailed picture is considered in \cite{Kapusta}, where
coupled rate equations for many hadron species 
are solved in the dynamical background of
2 + 1 dimensional hydrodynamics.});
(v) chemical equilibrium at $T_{chem}, \, \mu_{chem}$;
(vi) detailed balance and unitarity.
Items (i - v) are relaxed in transport codes, which also attempt
to include (vi). We insist here to arrive
at a qualitative and transparent understanding of an aspect of chemical freeze-out.
Items (i) and (ii) are related with the derivation of the employed
form of the momentum integrated Boltzmann equation, see \cite{Gondolo}
for details. Item (iii) refers to the fact that we consider a specific
expansion pattern of the fireball (which is assumed to contain a homogeneous 
matter distribution according to item (ii)) characterized by 
$V / \dot V = \bar \tau$ with $\bar \tau$ as expansion time scale, see
Appendix A. Item (v) fixes the initial conditions in agreement with
the thermo-statistical model \cite{hadro_chemistry}.

Strictly speaking, item (vi) refers to the balance 
$p + \bar p \leftrightarrow X + \bar X$ as the underlying Boltzmann equation
including a binary collision kernel
leads to our evolution equations (\ref{eq1}, \ref{eq2}). 
One may think, however, that both $X$ and $\bar X$
represent clusters of pions. Such states $X, \bar X$ are successfully
considered in
\cite{Cugnon} as two-meson doorway states coupling in turn to multi-pion states.
This approach, also known as minimal two-body model,
realizes the nearest threshold dominance and describes the $p \bar p$ annihilation 
data in the energy region relevant for our purposes.
It bridges to \cite{antip_puzzle_Shuryak,antip_puzzle_Rapp},
where such multi-pion collisions were considered as key to resolve
the antiproton puzzle together with chemical off-equilibrium
in the pion component. As further possible ingredients one may consider
the role of baryon excitations (as sources for further annihilations
and regenerations) and feeding of the ground state baryons
eventually observed, as done in \cite{Kapusta} for RHIC energy.

\section{Antibaryons in Big Bang}

\begin{figure}[h]  %
\includegraphics[width=0.68\textwidth]{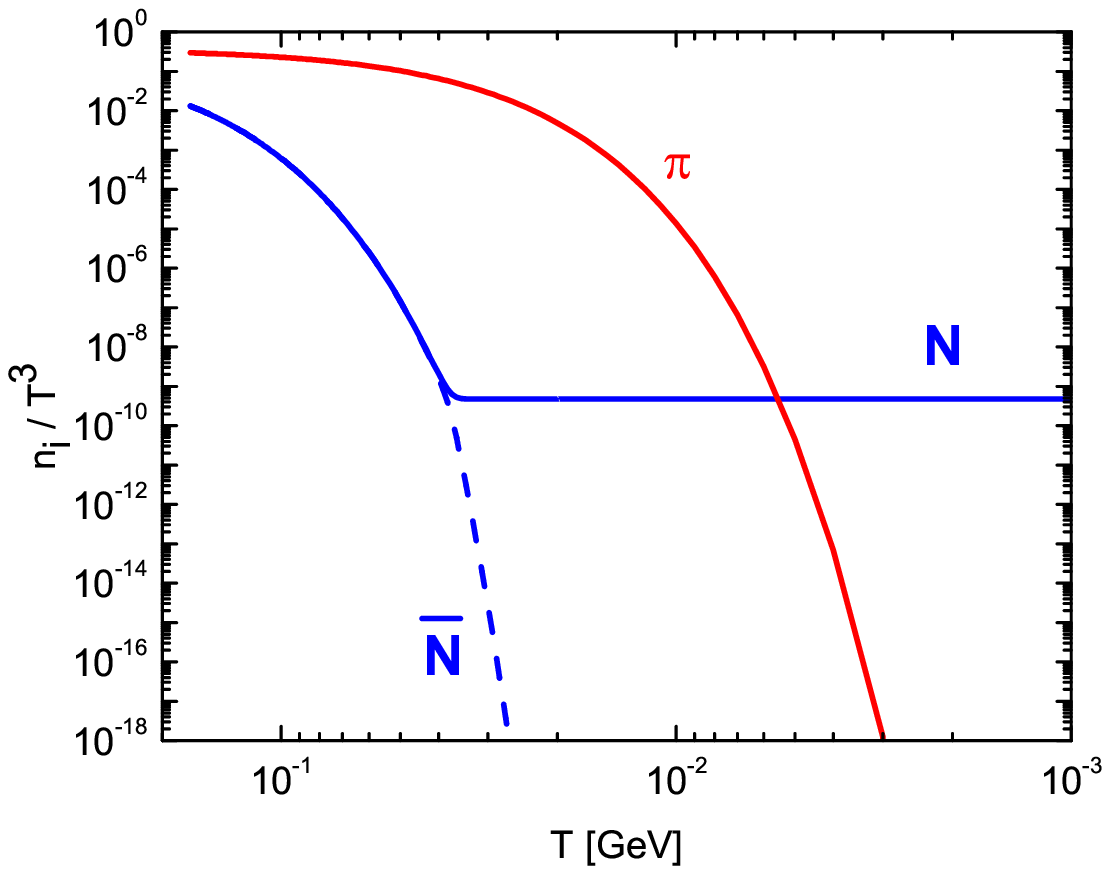} %
\includegraphics[width=0.68\textwidth]{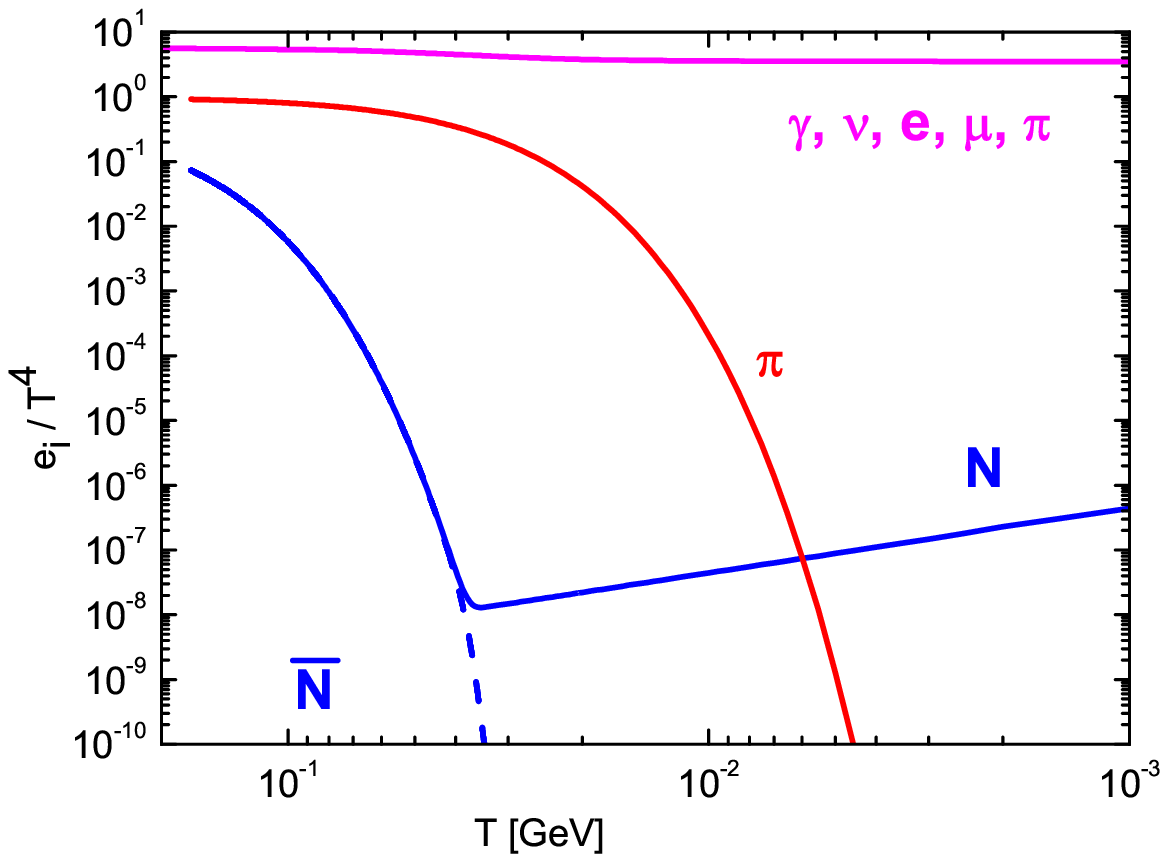} %
\caption{\label{fig5} (Color online)
Evolution of scaled hadron densities 
$n_i /T^3 = Y_i s / T^3 \approx Y_i^{eq} s / T^3$ 
(top panel) and related energy densities
(bottom panel; the top curve is for the electro-weak contribution
($\gamma$, $\nu$'s, $e$, $\mu$) + pions ($\pi$); 
the pion component is also depicted separately). 
Instead of $p$ and $\bar p$, the densities of nucleons ($N$, $n_+$) and
antinucleons ($\bar N$, $n_-$) are exhibited. For $\eta = 10^{-10}$.}
\end{figure}

Due to the numerically large
value of $\Lambda_{(2)}$, the evolution of $n_\pm$ follows closely the equilibrium
values $n_\pm^{eq}$ for a long time. In other words, $Y_\pm$ are dragged by
$Y_\pm^{eq}$ as evidenced by solving numerically Eqs.~(2.1, 2.2). In the temperature region
$T > 1$ MeV, $Y_\pm \approx Y_\pm^{eq}$ represent highly accurate solutions of the
evolution equations. Freeze-out of the antinucleon annihilation
happens at temperatures being considerably lower than the MeV scale \cite{Turner} thus
yielding an exceedingly small antibaryon density
in the assumed homogeneous scenario (for inhomogeneous scenarios cf.\ \cite{Jedamzik}).
Indeed, using $n_\pm \approx n_\pm^{eq} = s Y_{eq} \exp\{\pm \mu / T\}$,
one finds a rapid dropping of the scaled antibaryon density, represented by $n_-/T^3$,
with decreasing temperature, see top panel of Fig.~\ref{fig5}
($\mu$ is determined by Eq.~(4.1) below). For the given small surplus of baryons,
the scaled density $n_+/T^3$ follows closely $n_-/T^3$ until 40 MeV;
on the exhibited scale in Fig.~\ref{fig5} the difference of $n_+/T^3$ to $n_-/T^3$ is not
visible. Below that temperature of 40 MeV, however, $n_+/T^3$ stays approximately
constant at $\eta s/T^3$; the still continuing annihilations diminish
$n_+/T^3$ only marginally
\footnote{
Given $\Lambda_{(2)} \sim {\cal O}(10^{21})$, only for hypothetical values of
$\eta < 10^{-16}$ the antinucleon contribution
could be sizeable due to chemical freeze-out at $T \stackrel{<}{\sim} 30$ MeV,
as can be found in solving numerically Eqs.~(2.1, 2.2).}
since $n_-/T^3$ became small. The turn of $n_+ /T^3$ from dropping to the
approximately constant value near $\eta s/T^3$ at temperatures of 40 MeV is
entirely determined by the value of $\eta$ \cite{BKBluhm} 
which quantifies the baryon surplus, encoding also the chemical potential. 

\begin{figure}[h]  
\includegraphics[width=0.67\textwidth]{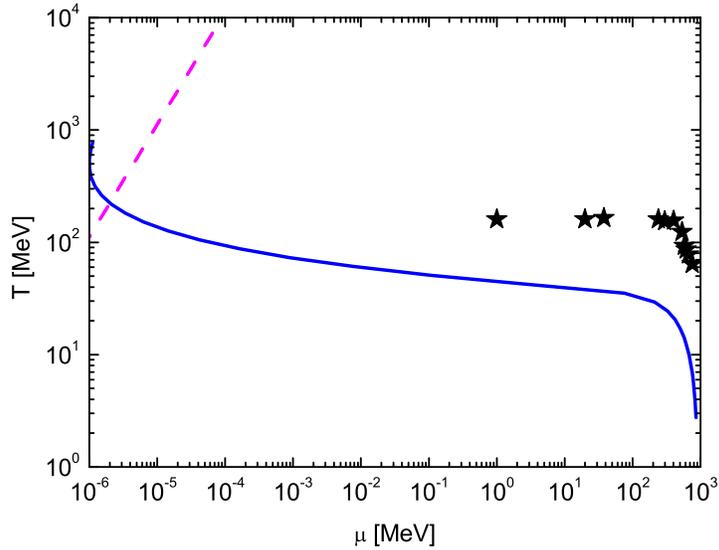} %
\vskip -6mm
\caption{\label{fig6} (Color online)
Adiabatic path of cosmic matter in the phase diagram of strongly
interacting matter for $\eta = 10^{-10}$ (solid curve for confined matter,
$h_{eff} = 10$).
The straight upper section (dashed curve) is for an approximation of deconfined matter.
(The crossing does not necessarily imply phase equilibrium. 
For the given approximations, the turn
from deconfined to confined adiabatic paths is by a Maxwell like construction
with mixed phase; these path sections are not displayed.)   
The asterisks depict chemical freeze-out points from \cite{hadro_chemistry}
(first quotation, table 2-upper part, and LHC estimate mentioned in
text there).}
\end{figure}

In the cosmic evolution, say after confinement, $\eta \sim 10^{-10}$ is
presumably realized, governed by the observed ratio of baryons to photons 
being $6.12 \times 10^{-10}$  
($\Lambda$CDM 3-year WMAP-only data \cite{WMAP}) and relying on adiabaticity. 
This ratio also determines the adiabatic path of cosmic matter
after confinement until the onset of primordial nucleo-synthesis
at $T \lesssim 1$ MeV, see Fig.~\ref{fig6} 
which is based on baryon conservation expressed by 
\begin{equation}
\mu = T\, \mbox{arsh} \left( \eta \frac{2 \pi^4}{45} 
\frac{h_{eff}}{2} \frac{1}{x^2 K_2(x)} \right).
\label{eq.3.1}
\end{equation}
Strictly speaking, this equation applies for $h_{eff} = const$
and for situations where nucleons are essentially the carriers of
baryon charge. Accordingly,
the cosmic baryo-chemical potential $\mu$ evolves from
$10^{-6}$ MeV at $T = 170$ MeV towards $m_N \sim 938$ MeV prior
to nucleo-synthesis at $T < 1$ MeV \cite{BKBluhm,Rafelski}.
Features of Eq.~(\ref{eq.3.1}) are:
(i) $min (\mu) = {\cal O} (m_N \eta)$,
(ii) $max (\mu) = m_N$,
(iii) crossing the $\mu = T$ line at
$T \approx m_N/(- \log \eta + \cdots)$ 
(this is the point where the difference of $n_+$ and $n_-$ becomes
large with dropping temperature: $n_+/T^3$ then stays approximately constant at
$\eta (2 \pi^2/45) h_{eff}$, 
while $n_-/T^3$ continues dropping
exponentially, see Fig.~\ref{fig5} and \cite{BKBluhm}),
(iv) before the region $\mu \sim T$ is reached, the
temperature drops as $T \propto m_N /(\log(\mu/m_N \eta) + \cdots)$.
The $\cdots$ in items (iii) and (iv) are for subleading terms.
The variation of $h_{eff}$ with temperature, according to the resonance gas
model, pulls down the isentropic curve (solid line) for $\mu < 1$ keV so that the
crossing with the dashed line occurs at a temperature of 165 MeV
(instead of 225 MeV). For $\mu > 1$ keV, the results of the resonance gas
model with adiabaticity and baryon conservation are on top of the
solid line.     
 
For an orientation,
in Fig.~\ref{fig6} also chemical freeze-out points
in relativistic heavy-ion collisions
from the analysis in \cite{hadro_chemistry} (first quotation)
are exhibited. At the freeze-out temperature of about 160 MeV
in Little Bangs at RHIC 
the baryo-chemical potential
is about $10^7$ times larger than in Big Bang. Otherwise,
when the baryo-chemical potential values of Little Bangs at RHIC
is achieved, the temperature in Big Bang is about 40 MeV
(see also \cite{PBM_Wambach}), being surprisingly high.

During the evolution of matter after electro-weak
symmetry breaking at $T \sim {\cal O} (100$ GeV) down to confinement
at $T_c \sim {\cal O} (200$ MeV), the strongly interacting matter
dominates by far the pressure, the energy density and the entropy density
\cite{Hindmarsh}. The masses of carriers 
of baryon charge change in the confinement transition.
Below 160 MeV the energy density is dominated by electro-weak
matter for a long time including primordial nucleo-synthesis, 
see bottom panel of Fig.~\ref{fig5}.
Pions are exceptional, as they are sizeable in number and energy density
contribution down to 5 MeV. The bottom panel of Fig.~\ref{fig5}
evidences also the slow relative increase of the scaled baryonic energy density;
much later (before ''recombination'') it will exceed the electro-weak
matter thus turning the radiation universe into a matter dominated universe.

\section{Summary}

In summary we contrast the baryonic antimatter evolution in Little Bangs
(i.e.\ bulk matter of fireballs created in relativistic heavy-ion collisions) and Big Bang
from confinement
towards the onset of primordial nucleo-synthesis.
To expose the similarities and differences we focus on antiprotons in
heavy-ion collisions and nucleons in the cosmic evolution
within a schematic kinetic description neglecting the explicit coupling to
other hadron states. 
The vast differences of Little Bangs and Big Bang are impressively
described by the huge difference of the dimensionless parameters $\Lambda_{(\xi)}$
governing the chemical freeze-out dynamics. In addition, the scaled
net baryon densities encoded in $\eta$ are also drastically different.

In Little Bangs, the ratio $\bar p / p$
drops only by a tiny amount from the established chemical freeze-out
temperature (which is defined by a multitude of other hadron abundances)
until kinetic freeze-out for RHIC conditions, where antiprotons
appear in a sizeable fraction. For a larger asymmetry of antiprotons
to protons (as realized for SPS conditions)
the ratio $\bar p / p$ changes still by a sufficiently small
amount to maintain the consistency of the thermo-statistical
model for hadron chemistry. The thermo-statistical model is an important tool for mapping
out the phase diagram of strongly interacting matter in the $T - \mu$
plane. In so far, it is important that its features conform
with a detailed description of heavy-ion collisions.  

In homogeneous scenarios for the Big Bang,
the chemical equilibrium is maintained for a long time
thus resulting in an exceedingly small fraction of primordial antibaryons \cite{Turner}.
An interesting point is the adiabatic path of cosmic baryons through the phase diagram
of strongly interacting matter and the related temperature of about
40 MeV (determined by $\eta$)
during the orders-of-magnitude change of the chemical potential
turning cosmic baryon matter from $\mu / T \ll 1$ into
$\mu / T \gg 1$. 

\acknowledgements
The work is supported by BMBF 06DR136, GSI and EU-I3HP.
The authors gratefully acknowledge stimulating discussions
with E. Grosse.

\appendix
\section{Momentum-integrated Boltzmann equation}

Let be $L_i N_i = C_i$ the Boltzmann transport equation for particles of species $i$,
where $L_i$ is the covariant Liouville operator,
$L_i = p^\mu_i \frac{\partial }{\partial x^\mu} 
- \Gamma^\mu_{\alpha \beta} p^\alpha_i p^\beta_i \frac{\partial }{\partial p^\mu_i}$ 
with affine connection
$\Gamma^\mu_{\alpha \beta}$ and Greek indices running from 0 - 3
(Einstein's sum convention applies only for them). $C_i$ may be a general
source term modifying the free-stream of the distribution function
$N_i (x, p)$ by collisions and feeding by decays. A consequence of the structure
of the Liouville operator is
${N_i^\alpha}_{; \alpha} = \int L_i N_i \Pi_i$ \cite{Stewart,Wagoner}
and thus with the Boltzmann equation one gets ${N_i^\alpha}_{; \alpha}= \int C_i \Pi_i$,
where the coordinate-independent momentum element for particle of species $i$ is
in the notation of \cite{Wagoner} 
$\Pi_i = d_i (2\pi)^{-3} \sqrt{-g} \frac{d^3 p_i}{\vert p^0_i\vert}$ with the fundamental
determinant $-g$, to be built of the metric tensor $g_{\mu \nu}$, and particle degeneracy
$d_i$; the semicolon stands for the coordinate-covariant derivative. 
The particle momenta $p_i^\mu$ are normalized to rest masses $m_i$
as $g_{\mu \nu} p^\mu_i p^\nu_i = - m_i^2$.
The particle current is defined by $N_i^\alpha \equiv \int N_i p_i^\alpha \Pi_i$.
For an observer moving with four-velocity $v_\alpha$ the corresponding particle density
is $n_i = - v_\alpha N_i^\alpha$. 

We consider here bulk matter in sufficiently small volume elements where
matter looks isotropically and homogeneously. Moreover, the matter is provided
to be in collisional (i.e., thermal) equilibrium so that a common flow field
$u^\alpha (x)$ can be attributed to all particle species. 
This means, $N_i^\alpha = n_i u^\alpha$ \cite{Zimdahl}.
Synchronizing the observer velocity with the flow by introducing a comoving
coordinate system where $u^\alpha = g^\alpha_0 \equiv \delta^\alpha_0$, the
l.h.s.\ of the balance equation for the density $n_i$,
$(n_i u^\alpha)_{; \alpha} = \int C_i \Pi_i$, becomes 
$\frac{1}{\sqrt{-g}} (\sqrt{-g} n_i g^\alpha_0)_{, \alpha} =
 \frac{1}{\sqrt{-g}} (\sqrt{-g} n_i)\dot{}$, 
where the dot means time derivative
in the respective comoving coordinate system. Introducing with the reasoning
of \cite{Wagoner} a comoving three-volume $V$, defined, e.g., by $(n_{con} V)\dot{} = 0$
for a conserved charge density $n_{con}$ or by entropy conservation, $(s V)\dot{} = 0$,
the l.h.s.\ of the considered balance equation may be written as
$\frac1V (n_i V)\dot{}$.

Binary elastic collisions, $i + i' \leftrightarrow i'' + i'''$, 
cancel out the r.h.s.\ of the balance equation as they represent collisional invariants.
For the moment being, we focus on binary annihilation processes, 
$i + \bar i \leftrightarrow X + \bar X$ \cite{Kolb_Turner}. In the absence of Bose
condensation or Fermi degeneracy effects and with the assumption that species
$X$ and $\bar X$ have equilibrium distributions, the collision term 
$\int C_i \Pi_i$ can be evaluated
(see \cite{Kolb_Turner,Gondolo} for details) to result in the master type equation
\begin{equation} \label{rate_equation}
\dot{n_i} + n_i \frac{\dot{V}}{V} = - \langle \sigma v \rangle
(n_i n_{\bar i} - n_i^{eq} n_{\bar i}^{eq}) 
\end{equation}
and an analog equation for $\bar i$ by the replacement $i \to \bar i$.
For Robertson-Walker-Friedmann cosmology, where strict isotropy and homogeneity
is applied, this is an often derived and applied equation, see 
\cite{Kolb_Turner,Gondolo,Turner,Griest}, and is referred to as momentum-integrated
Boltzmann equation.
In heavy-ion collisions such an equation, supplemented by decay terms, is
used too, for instance in \cite{Kapusta}.
In order to apply the above
arguments, however, special flow symmetries must be required. We consider examples below.
(Other approaches \cite{HW_BK} to rate equations use a spatial average 
over the expanding fireball, where
a Lorentz factor for transforming the time coordinate
to the external coordinate system occurs additionally.)

We follow further the arguments in \cite{Kolb_Turner,Gondolo} and reformulate the
rate equation (\ref{rate_equation}), instead as time evolution equation with
respect to comoving or proper observer time, as evolution equation with respect to
temperature $T$: Define the yields $Y_i = n_i / s$, 
use the definition of the comoving volume $V$ by $(s V)\dot{} = 0$,
and introduce the new variable $x = m_N / T$
to arrive at
\begin{equation} \label{rate_eq}
\frac{d Y_i}{d x} = - \langle \sigma v \rangle 
\left(Y_i Y_{\bar i} - Y_i^{eq} Y_{\bar i}^{eq}\right)
\frac{m_N^3}{x^4} \frac{3 V}{\dot{V}}
\frac{2 \pi^2}{45} h_{eff} \left(1 + \frac{\dot{h}_{eff}}{h_{eff}} 
\frac{T}{3 \dot{T}} \right). 
\end{equation}   
Clearly, further evolution equations are still needed describing the expansion
dynamics of the system under consideration. For our schematic discussion
in heavy-ion collisions we put $V / \dot{V} = \bar \tau$ with $\bar \tau$ as
characteristic expansion time. If the flow pattern is assumed to be specific,
the relevant time scales are henceforth prescribed.
Examples for simple flow patterns are the four-velocities
(i) $\bar u^\mu = (\mbox{ch} \hat\eta, 0, 0, \mbox{sh} \hat\eta)$,
(ii) $\bar u^\mu = (\mbox{ch} \hat\eta, \mbox{sh} \hat\eta \cos \phi, \mbox{sh} \hat\eta \sin \phi, 0)$,
(iii) $\bar u^\mu = (\mbox{ch} \hat\eta, \mbox{sh} \hat\eta \sin \theta \cos \phi, 
\mbox{sh} \hat\eta \sin \theta \sin \phi, \mbox{sh} \hat\eta \cos \theta)$
with $\hat\eta = \frac12 \log ((\bar t + \zeta)/(\bar t - \zeta))$, for 
(i) $\zeta = \bar z$, 
(ii) $\zeta = \sqrt{\bar x^2 + \bar y^2}$ and
(iii) $\zeta = \sqrt{\bar x^2 + \bar y^2 + \bar z^2}$ describing 
(i) purely longitudinal (Bjorken), 
(ii) axial-symmetric transverse und 
(iii) spherical expansion yielding with $\tau = \sqrt{\bar t^2 - \zeta^2}$
for $(1/V) (d V / d \tau)$
(i) $1/\tau$, (ii) $2/\tau$, and (iii) $3/\tau$
(here, $\bar t, \bar x, \bar y, \bar z$ are Cartesian coordinates in Minkowski space-time where
$\bar u^\alpha$ refers to; $\theta$ and $\phi$ are usual cylinder or polar coordinates).
The approximation leading to Eqs.~(2.1, 2.2) consists in the replacement of the
latter ratios by the inverse of a characteristic time scale, $1 / \bar \tau$.
Executing the transformation to the comoving coordinates
via $u^\alpha = \frac{\partial x^\alpha (\bar x)}{\partial \bar x^\mu} \bar u^\mu$
yields $u^\alpha = g^\alpha_0$; the line elements in the
comoving coordinates read 
(i) $ds^2 = -d \tau^2 + dx^2 + dy^2 + \tau^2 d\hat\eta^2$,
(ii) $ds^2 = - d \tau^2 + \tau^2 d\hat\eta^2 + \tau^2 \mbox{sh}^2 \hat\eta \, d\phi^2 + dz^2$,
(iii) $ds^2 = - d \tau^2 + \tau^2 d\hat\eta^2 + \tau^2 \mbox{sh}^2 \hat\eta \, d\theta^2
+ \tau^2 \mbox{sh}^2 \hat\eta \, \sin^2 \theta \, d\phi^2$ 
representing time-orthogonal (Gaussian) coordinates, as required to arrive
at Eq.~(\ref{rate_equation}).

In Robertson-Walker-Friedmann cosmology with the line element
$ds^2 = -dt^2 + R^2(t) d \vec x^{\, 2}$ the comoving velocity is as mentioned above;
$R(t)$ is the scale factor.
The Einstein equations govern the dynamics. For conformal flat three-space and without 
cosmological constant one has \cite{Bernstein} 
$\dot{V}/ V = 3 \dot{R} / R = 3 \sqrt{\frac{8 \pi}{3} G_N e}$ with
Newton's constant $G_N = M_{Pl}^{-2}$ and the energy density 
$e = \frac{\pi^2}{30} g_{eff} T^4$. Combining these quantities
appropriately one gets Eqs.~(2.1, 2.2) with (2.4) from (\ref{rate_eq})
as compact notation of the evolution
equations for protons or nucleons ($i \to +$) 
and antiprotons or antinucleons ($\bar i \to -$).

\newpage

\end{document}